\documentclass{article}
\usepackage{ijcai26}

\usepackage{times}
\usepackage{soul}
\usepackage{url}
\usepackage[hidelinks]{hyperref}
\usepackage[utf8]{inputenc}
\usepackage[small]{caption}
\usepackage{graphicx}
\usepackage{amsmath}
\usepackage{amsthm}
\usepackage{booktabs}
\usepackage{algorithm}
\usepackage{algorithmic}
\usepackage[switch]{lineno}

\title{Accessible, but Not Adopted: Increasing LLM Adoption among First-generation, Low-income (FGLI) College Students beyond Expanding Access}

\author{
    Hyungsik Kim
    \affiliations
    Harvard University
    \emails
    hyungsik.kim@hotmail.com
}

\begin{document}

\maketitle

\begin{abstract}
Large language models (LLMs) are increasingly positioned as a force to empower underserved communities, and significant efforts are being made to expand access. Yet, access alone does not equate to meaningful adoption. First, even if a system is accessible, it won't be adopted if users are not willing to adopt it. Second, even if an LLM system is superficially adopted, the heterogeneity of LLM tools means that LLM adoption can be further deepened. Closing this access-adoption gap is critical to ensuring that the full social potential of LLM is not only accessible but fully realised. Drawing on 61 interviews (15 long-form semi-structured interviews with first-generation, low-income college (FGLI) students, 3 non-FGLI students, 3 FGLI program directors, and 40 intercept interviews), this paper examines the access-adoption gap in first-generation, low-income student communities. This paper a) finds that while FGLI students have adopted LLM systems, their depth of LLM tool usage is limited to chatbots (e.g., ChatGPT or Claude) for narrow use cases, and b) identifies barriers limiting their willingness to learn and use (low perceived value, under-estimated self-efficacy, unclear starting point, low peer exposure, and resource constraints). Then, from these findings, the paper derives the four design principles to design a system or an intervention aimed at closing the access-adoption gap in LLM adoption by FGLI students. In doing so, the paper contributes to the field by a) examining the LLM access-adoption gap in the FGLI student community, and b) reframing LLM adoption as a depth gradient across four modes of LLM tool use: basic chatbot interfaces, tool-augmented prebuilt interfaces, agentic development interfaces, and programmatic integration.

\end{abstract}

\section{Introduction}
As large language models (LLMs) have exhibited potential as an equalising force for underserved communities, great efforts have been made to improve access both in how LLM systems are designed and distributed. 

Yet, access alone does not equate to meaningful adoption of LLM systems. First, improved access does not necessarily lead to adoption. If users are not willing to adopt a system due to internal or external factors, the system won't be adopted even if it is accessible. Second, even if an LLM system is superficially adopted, the heterogeneity of LLM tools means that LLM adoption can be further deepened. Furthermore, in underserved communities, this gap can be aggravated by structural challenges these communities face.

Closing this access-adoption gap is critical to ensuring that the full social potential of LLMs is not only accessible but fully realised in underserved communities, who face not only access barriers to LLMs, but also barriers to deep adoption due to limited social networks, time poverty, and resource constraints.

While existing work on technology adoption has explored this access-adoption gap, there is limited work in how such models apply specifically to the context of LLM adoption for underserved communities. 

Therefore, this paper seeks to answer three key research questions in the context of LLM adoption by first-generation, low-income (FGLI) students:
\begin{itemize}
    \item \textbf{RQ1}: What is the current depth of generative language model adoption among FGLI students?
    \item \textbf{RQ2}: What are the barriers that prevent adoption by FGLI students?
    \item \textbf{Q3}: What are the design principles for designing a system or an intervention aimed at closing access-adoption gap in LLM adoption by FGLI students?
\end{itemize}

To answer the questions, qualitative methods were used, including semi-structured long-form interviews with 15 first-generation, low-income (FGLI) college students, 3 FGLI program directors in colleges, short interviews with 3 non-FGLI students, and 40 short intercept interviews with students across multiple colleges. 

This paper characterises the depth of LLM use in terms of four modes: basic chatbot interfaces (e.g., conversational use of ChatGPT, Claude, or Gemini), tool-augmented prebuilt interfaces (e.g., deep research, multi-modal generations), agentic development interfaces (e.g., Claude Code, Codex), and programmatic integration through LLM APIs and SDKs. 

This paper demonstrates the access-adoption gap by a) showing that while FGLI students have adopted LLM systems, their depth of LLM tool usage is limited to chatbots (e.g., ChatGPT or Claude) for narrow use cases, and b) identifying barriers that are limiting their willingness to learn and use (intent to learn and intent to use). Then, from these findings, the paper derives the four design principles to design a system or an intervention aimed at closing the access-adoption gap in LLM adoption by FGLI students.

\section{Related Work}
\subsection{Technology adoption}
The following perspectives on technology adoption provide a conceptual foundation for identifying barriers to language model adoption facing under-resourced college students. 

The Technology Acceptance Model (TAM) \cite{davis_perceived_1989} builds on the theory of planned behaviour \cite{Ajzen1985}, which posits that behaviours are driven by intention, identifying \textit{perceived usefulness} and \textit{perceived ease of use} as the drivers of the \textit{intent to use} technology. Venkatesh et al. extend TAM, in their Unified Theory of Acceptance and Use of Technology (UTAUT) by adding external drivers of intent to use, including \textit{social influence} and \textit{facilitating conditions} \cite{venkatesh:utaut}. 

Kim and Gajos further deepen the model by distinguishing between \textit{intent to use} and \textit{intent to learn}, underscoring the significance of additional effort invested to learning the new technological system even before using it \cite{kim_acceptance_2016}. 

The theory of self-efficacy offers an additional lens to understanding the intention: people adopt new technologies when they are confident in their capacity to use them effectively. Bandura identifies structured mastery experiences, vicarious modelling, verbal persuasion, and psychological safety \cite{bandura_self-efficacy_2012}.

\subsection{AI literacy and fluency}
The literature on AI competency provides more multidimensional perspectives on adoption that extend beyond the binary view. 

Long and Magerko define AI literacy as a multi-dimensional competency set (recognising AI, understanding how it works, critically evaluating, using and applying, engaging ethically) \cite{long_what_2020}. Rogers and Carbonaro distinguish between \textit{AI literacy}, which describes conceptual understanding and \textit{AI fluency}, which describes the capacity to deploy AI in practice \cite{rogers_understanding_2025}. 

While these frameworks conceptualise adoption as a multidimensional set, they do not conceptualise it explicitly as a \textit{depth of tool use} (e.g., chatbot, tool-augmented use, agentic orchestration).

\subsection{FGLI students' LLM adoption}
Though the field focusing specifically on LLM adoption is emerging, there is a large body of work highlighting the structural inequalities in technology adoption by FGLI students. For example, Pearson et al. argue that barriers to persistence and retention in STEM education has led to FGLI students being underrepresented in STEM degree completion and identify key support elements (e.g., mentorship and structured support) that can improve STEM persistence \cite{pearson_systematic_2022}.

On LLM adoption specifically, there is growing evidence in grey literature for awareness and adoption gaps between continuing-generation students and FGLI students \cite{reeves2023edtech}. Arum et al. also empirically demonstrate that continuing-generation students are significantly more likely than first-generation students to be aware of ChatGPT \cite{arum_chatgpt_2025}.

\textbf{This paper} seeks to build on existing work by focusing on two areas where less effort has been made. First, it applies the intent-driven adoption frameworks to FGLI students' LLM adoption to identify key barriers. Second, when evaluating LLM adoption, it builds on a multidimensional perspective on AI fluency by conceptualising LLM adoption in terms of the depth of LLM tool usage rather than a binary threshold. 

\section{Method}
The findings were drawn from 15 semi-structured interviews and 40 intercept interviews with self-identified FGLI students 3 non-FGLI students, and 3 administrative staff members responsible for FGLI support programs. 

\textbf{Participants}. 15 FGLI participants (mainly freshmen/sophomores) for semi-structured interviews were recruited at Harvard University through WhatsApp group chats, mailing lists targetting FGLI-identified students, and residential house mailing lists. 3 non-FGLI students were recruited through the personal network. 3 FGLI program directors from 3 different research universities were recruited directly via email. For intercept interviews, 40 students were recruited on the spot in the cafeteria at Tufts University (10), the entrance of Borough of Manhattan Community College (BMCC) (10), and in the library and cafeteria at New York City College of Technology (20).

\textbf{Data collection}. Three types of interview protocols were used. 
\begin{enumerate}
    \item For semi-structured interviews with students, the protocol focused on questions about the depth of LLM tool use (in terms of types of LLM tools used), reasons for using LLMs, perception of LLM tools, and barriers to deeper use of LLM tools.
    \item For intercept interviews with students, a shorter version of 1) was used with particular focus on depth of LLM tool use and barriers to deeper use of LLM tools.
    \item For semi-structured interviews with FGLI program directors, the protocol focused on structural barriers facing FGLI students.
\end{enumerate}
Five interviews were conducted by project collaborators using the shared interview protocol; the remaining interviews were conducted by the author. 

\textbf{Adoption depth framework}. The depth of LLM adoption was characterised in this paper using four modes that broadly reflect increasing implementation responsibility and technical control:
\begin{enumerate}
    \item Basic chatbot interface: a ready-made conversational interface for discrete, user-directed tasks without additional tools or direct system configuration (e.g., conversational use of ChatGPT, Claude or Gemini).
    \item Tool-augmented prebuilt interface: built-in capabilities beyond basic conversation (e.g., deep research, multimodal generation).
    \item Agentic development interface: a prebuilt interface that uses agentic planning to determines and executes intermediate actions to create or modify a software artefact (e.g., Claude Code, Codex).
    \item Programmatic integration: direct use of LLM APIs or custom code to embed or orchestrate model
capabilities within an application or system.
\end{enumerate}

\textbf{Analysis}. Interview data was analysed using inductive and deductive approaches informed by existing theoretical concepts. First, key themes based on codes and categories that emerged from the existing theories and models (e.g., performance expectancy, self-efficacy, intention to learn) have been identified. Then, new themes were identified inductively based on the interview data to further enrich the initial set of themes (e.g., career relevance, time poverty). These themes were classified to answer the research questions. When participants are quoted, to protect their identities, they are identified using anonymised codes (e.g., S01 for students and A01 for FGLI program directors).

Beyond the interview data, we derive design principles for closing the access-adoption gap in the FGLI student community. Each design principle is mapped directly to the themes from the thematic analysis of the interview data. These principles are inferences from the empirical findings rather than separately tested claims.

\section{Results}
In this section, we present our primary findings.
\subsection{RQ1: Depth of LLM adoption}
Three main patterns characterise the current depth of LLM adoption among FGLI students. The primary finding is that while the vast majority of FGLI students have "adopted" LLM tools, their depth of usage remains shallow and narrow.

\textbf{F1.1. Shallow adoption limited to chatbots.} While the vast majority of the participants indicated that they use LLMs, (with exception of 2 participants who indicated the use of vibe coding tools such as Claude Code), none of 55 FGLI participants reported meaningful use of LLMs beyond the chatbot interface (e.g., ChatGPT, Claude). Even when other tools such as vibe-coding platforms, agentic tools, and image/video generation were mentioned in follow-up questions, participants did not indicate that they use these LLM tools and even expressed a lack of awareness of the tools. 

\textbf{F1.2. Narrow LLM use cases for academic or career support}. The vast majority of the participants expressed the use of LLM tools for academic purposes (incl. concept explanation, "office hour" substitute, problem walkthroughs, research):
\begin{quote}
"Most often use it for learning concepts for like my homework. Certain times when I'm working on assignments and office hours won't be for quite a while or like over the weekend" (S13) 
\end{quote}
Other common uses included career support (e.g., resume review, interview preparation) and coding (e.g., syntax review, coding automation (on a chatbot)). 

\textbf{F1.3. Low LLM tool awareness besides chatbots}. FGLI students explicitly mentioned that they are behind their peers in their understanding of different available LLM tools:
\begin{quote}
"My knowledge of AI is very limited compared to my friends... I am not as knowledgeable about AI as I could be." (S13)
"I didn't realise that [Codex] is out there where you could have very limited coding knowledge and you could create like fully functional apps." (S12)
\end{quote}
The gap was even more stark in non-research universities outside Harvard, where FGLI students majoring in computer science were not aware of the coding agents (such as Claude Code or Codex).

\subsection{RQ2: Barriers to adoption}

\textbf{F2.1. Low perceived value (esp. career relevance)}. FGLI students could not articulate a concrete value for themselves. This was most strongly and frequently observed in students in non-CS fields such as pre-med, law, social services, nursing, and English literature, where LLMs were repeatedly perceived as tools that are useful only for "tech people" rather than to students' own fields.
\begin{quote}
"If I'm being honest. I don't really see how useful it would be for my future career... AI is more useful for careers that are tech oriented." (S15, a sociology major)
"I did not know it was relevant for my career until I saw all the full-time job posts that talk about AI." (S05)
\end{quote}
Even students with interest in computer science indicated that they struggle to find projects where LLMs could be helpful:
\begin{quote}
"I'm not really sure what sorts of projects I could make with them other than essay help or writing purposes" (S14)
\end{quote}
Consistent with the technology adoption literature, the low performance expectancy (often framed as low career relevance) inhibits deeper investment in increasing the depth of adoption.

\textbf{F2.2. Under-estimated self-efficacy}. Almost all of the participants, regardless of their majors, expressed low confidence in their ability to understand LLM tools:
\begin{quote}
"I'm not sure if I'm going to be able to use every tool as effectively as like how I'm using it now and it might just be like a big waste of time of trying to like oh let me use this." (S12)
\end{quote}
What is worth noting is that a limited understanding of LLM tools (F1.3.) has led to over-estimation of the knowledge and skills needed to use LLM tools (e.g., understanding mathematics behind an LLM is not required for everyday use of coding agents), which in turn leads to under-estimation of their self-efficacy:
\begin{quote}
"I am not a tech person. I don't understand all the maths behind AIs." (S08)
\end{quote}
In Bandura's framework, the structural absence of mastery experiences (worsened by over-estimation of needed skills) produces pre-emptive disengagement from the use LLM tools.

\textbf{F2.3. Unclear starting point.} Many students frequently expressed feeling overwhelmed by not knowing where to begin, highlighting fragmented learning resources, (ironically) \textit{increased access} to similar LLM tools availability, and rapid evolution in the field. This is directly linked to a reduced willingness to learn (\textit{intent to learn}):
\begin{quote}
"There's difference between like Claude and ChatGPT and like how do they give answers like that? (...) I'm still confused (...) it takes a lot of effort which I guess ties in with the high effort. It just takes a lot of time to understand." (S12)
\end{quote}
For FGLI students, this challenge is aggravated by a) a lack of social networks or mentors who can provide structure to guide their learning.

When asked about how they would best improve their understanding of LLM tools, many cited scaffolded learning ("step-by-step from fundamentals" (S13)) as the most helpful support.

\textbf{F2.4. Low peer exposure}. When compared to the continuing-generation peers, the most notable difference in contextual factors was that FGLI students lack pre-college and peer ecosystems (e.g., national labs, tech-company camps, family in the tech sector (S16, S17, S18)) that create strong early AI exposure. Consistent with Bandura's vicarious modelling mechanism \cite{bandura_self-efficacy_2012}, and social influence in UTAUT \cite{venkatesh:utaut}, this low exposure works against deeper LLM adoption.

\textbf{F2.5. Resource constraints}. Furthermore, coming from under-resourced communities, FGLI students face financial and temporal constraints that moderate their willingness to invest in learning and adopting the LLM tools. These barriers directly affect the intent to adopt, but also intensify the effect of other contributing factors (F2.1, F2.2, F2.3, F2.4).

In the short term, this is linked to concerns about token and subscription costs.
\begin{quote}
"I gotta go work now. I don't have time to think about AI" (S30)
"The one time I did try [vibe coding] tokens were like a big issue. Like a hurdle that had to get around like oh I can't use [it] because I have a limited number of coins" (S13)
\end{quote}
In the long term, this is related to how FGLI students are forced to prioritise their time towards more high-income earning activities and other academic commitments:
\begin{quote}
"Getting a job that can give me financial independence is my primary concern." (S07)
\end{quote}
This pressure is reinforced by the sense of being behind their peers academically:
\begin{quote}
"Freshman year is kind of seen as a transition from high school to college… I just want to focus on my grades." (S10)
\end{quote}
\subsection{Q3. Design implications for closing the adoption gap}
In this sub-section, the design principles for systems or interventions aimed at improving LLM adoption in the FGLI student communities can be derived. This set of principles is not an exhaustive set for designing an intervention for the community. Instead, the goal is to address the access-adoption gap that is related to students' willingness to adopt LLM systems more deeply. It should be noted that each principle is derived from the primary findings and that they have not been separately tested.

\textbf{DP1. Demonstrate user-specific values.} As noted in F2.1., the majority of FGLI students did not perceive LLM tools to be valuable, leading to early dismissal of the tools despite their availability. A clear, tangible demonstration of concrete value to the users (particularly centred on career relevance) would address the issue of low performance expectancy directly. Furthermore, this perceived performance expectancy can be further increased if the value is demonstrated through vicarious modelling by their own peers (linked to F2.4).

\textbf{DP2. Clarify competency needs specific to the LLM tools.} Lack of awareness of different LLM tools (as noted in F1.3.) has resulted in subsequent over-estimation of the competencies needed to adopt them. Clarifying the required competency upfront can "de-mystify" inflated concerns around competency gaps that undermine students' self-efficacy.

\textbf{DP3. Scaffold learning from fundamentals to fluency}. Structured learning has been identified as the most helpful support for FGLI students to deepen LLM adoption. By addressing the feeling of being overwhelmed by uncertainty and fragmentation, the students' expected cost of learning would be reduced strengthening the \textit{intent to learn}.

\textbf{DP4. Proactively address resource barriers}. FGLI students often need to absorb the impact of the resource constraints, which are often binding. Proactively relieving these constraints or even simply addressing their concerns around the constraints (e.g., highlighting low time-commitment) where relevant can greatly enhance their intent to adopt LLM tools. 

\section{Discussion and limitations}
The empirical contribution of this paper is limited by three key factors. First, the sample is geographically concentrated on the US East Coast (specifically Boston and New York). The findings may not generalise to other regions. Second, the majority of the participants are freshmen or sophomore college students, and given that the analysis is cross-sectional, the findings cannot speak to how adoption trajectories evolve over time. Third, LLM tools are rapidly evolving every month, and so is their adoption. For example, the ChatGPT adoption rate in \cite{arum_chatgpt_2025} (published in 2025) is probably already outdated. 

Beyond its limitations, further areas of study include exploration of more concrete methods of evaluating the depth of LLM tool use that are still sensitive to the diversity of LLM tools, interaction between the factors driving the access-adoption gap and different LLM tools, and potential refinement/modification of the design principles based on empirical evidence.

\section{Conclusion}
Technology adoption is a process, not an event, and the findings demonstrate that it is particularly true for FGLI students' LLM system adoption in two ways. First, even after LLM systems become accessible, there must be an intent to learn and use the system for adoption to take place. Second, initial surface-level adoption can be further deepened through more advanced LLM tool uses. 

The findings demonstrate that for FGLI students, there are gaps on both fronts. To fully harness the potential of LLM systems to empower the FGLI student community, closing these gaps is critical.

\section*{Acknowledgements}
The author would like to thank Professor Julie Battilana for valuable guidance, and Richael Saka and Maya Ganesh for conducting selected participant interviews and for discussions that informed the development of the adoption-depth framework.

\bibliographystyle{named}
\bibliography{ijcai26}

\end{document}